\documentstyle[twocolumn,seceq,psfig,multicol]{jpsj}

\def\runtitle{Spin Diffusion in $\alpha$-VO(PO$_3$)$_2$}
\def\runauthor{Jun {\sc Kikuchi} \it et al.}

\title
{
Spin Diffusion in the $S=1/2$ Quasi One-Dimensional Antiferromagnet
$\alpha$-VO(PO$_3$)$_2$ via $^{31}$P NMR}

\author
{ 
Jun {\sc Kikuchi}\footnote{E-mail: kikuchi@ph.noda.sut.ac.jp},
Naohiko {\sc Kurata}, Kiyoichiro {\sc Motoya},\\
Touru {\sc Yamauchi}$^{1}$ and Yutaka {\sc Ueda}$^{1}$ }

\inst
{
Department of Physics, Faculty of Science and Technology, Science
University of Tokyo,\\
2641 Yamazaki, Noda, Chiba 278-8510\\
$^1$Institute for Solid State Physics, University of
Tokyo, 5-1-5 Kashiwano-ha, Kashiwa, Chiba 277-8581\\
}

\recdate
{Febrary 15, 2001}

\abst
{
Low-frequency spin dynamics in the $S=1/2$ antiferromagnetic
spin-chain compound $\alpha$-VO(PO$_3$)$_2$ has been studied by means
of $^{31}$P NMR. The nuclear spin-lattice relaxation rate $1/T_1$ at
the P site exhibits $H^{-1/2}$ dependence on the applied magnetic
field ($H$) at temperatures ($T$) well above the intrachain coupling
strength $J/k_{\rm B} = 3.50$ K indicating one-dimensional diffusive
spin dynamics.  The diffusive contribution to $1/T_1$ decreases on
cooling as electronic spins acquire short-range antiferromagnetic
correlations within the chain, and vanishes almost entirely around $T
\approx J/k_{\rm B}$.  This is accompanied by an apparent increase of
the spin-diffusion constant from the value expected in the classical
limit.  On the other hand, the field-independent part of $1/T_1$
increases with decreasing temperature, which may be a precursor for
the true long-range antiferromagnetic ordering found below $T_{\rm N}
= 1.93 \pm 0.01$ K. }

\kword {$\alpha$-VO(PO$_3$)$_2$, NMR, antiferromagnetic spin chain,
nuclear spin relaxation, spin diffusion}

\begin{document}
\sloppy
\maketitle

\section{Introduction}
\label{sec:intro}
There has been a continued interest in the dynamics of low-dimensional
quantum antiferromagnets at finite temperatures.  One of the issues which has
attracted renewed attention is a problem of spin diffusion in the
one-dimensional (1D) Heisenberg spin chain.  It is argued from
phenomenology that at high enough temperatures and at long times, the
spin autocorrelation function of the 1D Heisenberg spin chain has a
diffusive form
\begin{equation}
    \langle S_{i}(t) S_{i}(0)\rangle\propto t^{-1/2},
\label{eq:diffusion}
\end{equation}
    
\noindent
leading to divergence of the spectral density at low frequencies as
$\omega^{-1/2}$.  However, because the diffusive form
(\ref{eq:diffusion}) is not derived from the microscopic Hamiltonian
but is a consequence of the hydrodynamical assumption for the
spin-spin correlation,\cite{degennes63} the question on the existence of spin
diffusion in 1D spin chains has been studied intensively from both
theoretical and experimental viewpoints.  One of the best studied
theoretical models is the $S=1/2$ {\it XXZ} chain represented by the
Hamiltonian
\cite{sachdev94,narozhny96,fabricius98,naef98,narozhny98,zotos99}
\begin{equation}
    {\cal H} = J \sum_{i} (S_{i}^{x}S_{i+1}^{x} + S_{i}^{y}S_{i+1}^{y}
    +\Delta S_{i}^{z}S_{i+1}^{z}).
\label{eq:XXZ}
\end{equation}

\noindent
At the isotropic point ($\Delta =1$) and at low enough temperatures
($T\ll J/k_{\rm B}$), analytical expressions for the dynamical
susceptibility $\chi(q,\omega)$ have been derived for $q\approx 0$ and
$\pi$, and are shown to have no diffusive pole at low
frequencies.\cite{sachdev94} At modestly high, or much higher
temperatures compared with the intrachain coupling strength $J$, the
problem is still controversial.  Although the absence of diffusive
excitations seems to be settled for the {\it XXZ} chain with planar
anisotropy ($0\leq \Delta <1$), a definite answer for the absence (or
presence) of spin diffusion at the isotropic point has not yet been
given.

Diffusive behavior of the spin-spin correlation has been observed
experimentally in several 1D spin chains via the $\omega^{-1/2}$
resonance-frequency dependence of the nuclear spin-lattice relaxation
rate $1/T_1$ at elevated
temperatures.\cite{hone74,borsa74,barjhoux77,ajiro78,takigawa96-1,kikuchi97}
As to the $S=1/2$ chain, an analysis of the $\omega$ dependence of
$1/T_1$ at the Cu site in Sr$_2$CuO$_3$ based on the classical
spin-diffusion theory gives an unusually high value of the spin
diffusion constant compared with the classical
value.\cite{takigawa96-2} This suggests the absence of spin diffusion
in the low-temperature limit in consistency with the
field-theoretical\cite{sachdev94} and perturbative\cite{narozhny96}
approaches.  On the other hand, an evidence for diffusive spin
transport is found from the proton NMR measurements in 
CuCl$_2$$\cdot$2NC$_5$H$_5$ and Cu(C$_6$H$_5$COO)$_2$$\cdot$3H$_2$O at
temperatures well above the intrachain exchange
interactions.\cite{ajiro78} Therefore, there seems to be a gap between
the high- and low-temperature behavior of the spin transport in the
$S=1/2$ chains, necessitating further experiments on the dynamics,
especially on the temperature-dependent properties.

\begin{figure}[t]
\centerline{\psfig{figure=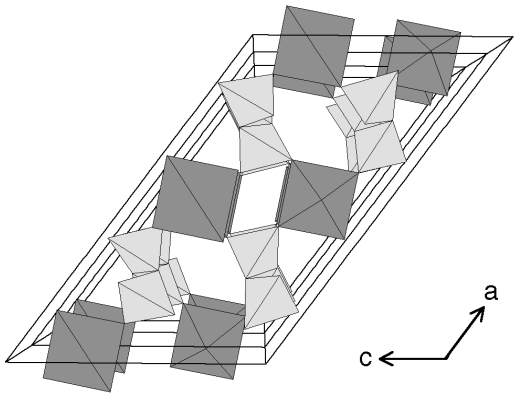,width=7.0cm,angle=0}}
\caption{Crystal structure of $\alpha$-VO(PO$_3$)$_2$ viewed from the
monoclinic $b$ axis.  The dark- and light-shaded polyhedra represent VO$_5$
pyramids and PO$_4$ tetrahedra, respectively.}
\label{fig:structure}
\end{figure}

Linear chains of a V$^{4+}$ ion in the compound $\alpha$-VO(PO$_3$)$_2$
may be a model system of an $S=1/2$ Heisenberg spin chain. 
$\alpha$-VO(PO$_3$)$_2$ belongs to the monoclinic space group $C2/c$
and has the room-temperature lattice parameters; $a = 15.140$ {\AA}, $b
= 4.195$ {\AA}, $c = 9.573$ {\AA} and $\beta =
120.54^{\circ}$.\cite{murashova94} In the $\alpha$-VO(PO$_3$)$_2$
structure (Fig.\ \ref{fig:structure}), VO$_5$ pyramids are stacked
along the $b$ axis to make up a linear chain of V atoms with the
nearest-neighbor V-V distance of 4.915 {\AA}.  The linear chains are
well separated by distorted PO$_4$ tetrahedra in the $a$ and $c$
directions, so that good one dimensionality in the $b$ direction is
expected.  In this paper, we report on the experimental study of the
low-frequency spin dynamics in $\alpha$-VO(PO$_3$)$_2$ via $^{31}$P
NMR. The relatively small intrachain coupling ($J/k_{\rm B} = 3.50$ K
estimated from the present susceptibility measurement) in
$\alpha$-VO(PO$_3$)$_2$ enables us an experimental access to a wide
range of temperatures not only $T \sim J/k_{\rm B}$ where short-range
antiferromagnetic correlations are important, but also $T \gg J/k_{\rm
B}$ where electronic spins behave almost paramagnetically.  Crossover
of the dynamics between the two temperature regimes can also be
elucidated by examining the temperature-dependent behavior of, for
example, the nuclear spin-lattice relaxation rate.  This type of the
experiment cannot be done in the canonical $S=1/2$ Heisenberg
antiferromagnetic spin-chain compound Sr$_2$CuO$_3$ which has a huge
intrachain exchange ($J/k_{\rm B} = 2200$ K)\cite{motoyama94} being
suitable for the study of low-temperature dynamics, and will give
complemental information for thorough understanding of low-energy spin
excitations in the $S=1/2$ Heisenberg spin chain.

\section{Experiments}
\label{sec:expts}
Polycrystalline samples of $\alpha$-VO(PO$_3$)$_2$ were prepared by a
solid-state reaction method.  Equimolar mixture of (VO)$_2$P$_2$O$_7$ and
P$_2$O$_5$ was fired in an evacuated silica tube at 750 $^\circ$C for
2 days and at 900 $^\circ$C for 2 days with the intermediate grinding. 
(VO)$_2$P$_2$O$_7$ was prepared as described in ref.\ \citen{kikuchi99}.  The
obtained samples were examined by the X-ray diffraction measurement
and were confirmed to be a single phase.  Magnetic susceptibility was
measured using a SQUID magnetometer (Quantum Design MPMS-5s) at 0.1 T.
NMR measurements were performed with a standard phase-coherent pulsed
spectrometer.  $^{31}$P NMR spectrum was taken by recording the
spin-echo signal with a Box-car averager at a fixed frequency while
sweeping the external magnetic field.  The nuclear spin-lattice
relaxation rate of $^{31}$P was measured by a saturation recovery method with a
single saturation rf pulse.  The measured nuclear-magnetization
recoveries were single exponential as expected for nuclei with spin
1/2, so that the nuclear spin-lattice relaxation time $T_1$ can
uniquely be determined as the time constant of the recovery curve.

\section{Results and Analysis}
\label{sec:results}
\subsection{Magnetic Susceptibility}
\begin{figure}[t]
\centerline{\psfig{figure=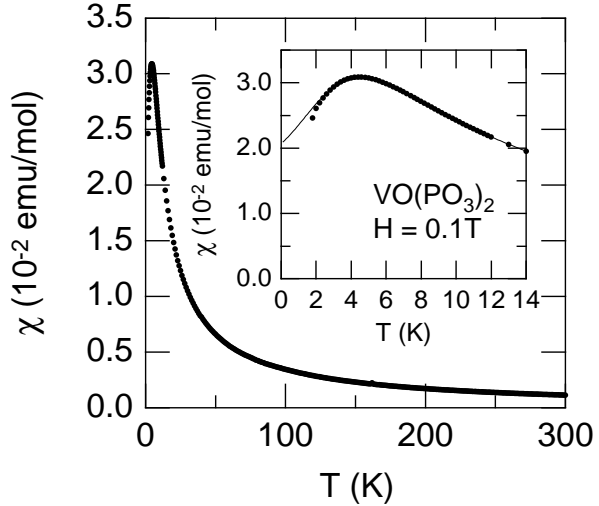,width=9.0cm,angle=0}}
\caption{Temperature dependence of the magnetic susceptibility of
$\alpha$-VO(PO$_3$)$_2$.  The inner core diamagnetic susceptibility
($-9.2 \times 10^{-5}$ emu/mol) was subtracted from the raw data.  The
inset is the expanded view at low temperatures.  The solid line in the
inset is a fit of the data to the susceptibility of an $S=1/2$
Heisenberg antiferromagnetic linear chain.}
\label{fig:suscept}
\end{figure}

Figure \ref{fig:suscept} shows the temperature ($T$) dependence of the bulk
magnetic susceptibility ($\chi$) of $\alpha$-VO(PO$_3$)$_2$.  While showing
Curie-Weiss-like behavior at high temperatures, the susceptibility
takes a rounded maximum around $T_{\rm max} \approx 4.4$ K as a
sign of low-dimensional nature of the exchange coupling.  The
susceptibility can nicely be fitted to the Bonner-Fisher curve for the
$S=1/2$ Heisenberg antiferromagnetic spin chain (HAFC).\cite{bonner64}
We estimated the intrachain coupling $J$ between V$^{4+}$ spins by
fitting the $T$ dependence of the bulk $\chi$ to the
formula;\cite{hatfield81}
\begin{eqnarray}
    \chi =&&\chi_{0}
    + \frac{N_{\rm A}\,g^2 \mu^2_{\rm B}}{k_{\rm B}T}
    \nonumber \\    &&\times 
    \frac{0.25 + 0.14995\,x +0.30094\,{x^2}}{1 + 1.9862\,x +
    0.68854\,{x^2} + 6.0626\,{x^3}}.
\end{eqnarray}

\noindent
Here $\chi_{0}$ is a $T$-independent part of $\chi$, $N_{\rm A}$ is
Avogadro's number, $g$ is Land\'{e}'s $g$ factor, $\mu_{\rm B}$ is
the Bohr magneton and $x=J/k_{\rm B}T$.  From the least-squares fit of
all the available data points (1.8 to 300 K), we obtained $\chi_{0}=
1.05(8) \times 10^{-5}$ emu/mol, $g$ = 1.978(1) and $J/k_{\rm B}$ =
3.50(1) K. The result of the fit is shown in the inset of Fig.\
\ref{fig:suscept}.  $\chi_{0}$ may be interpreted as the Van-Vleck orbital
paramagnetic susceptibility $\chi_{\rm VV}$, and the obtained value of
$\chi_{0}$ is in a reasonable range for $\chi_{\rm VV}$ of a V$^{4+}$ ion. 
The bulk $\chi$ deviates slightly from the Bonner-Fisher curve below
about 2.2 K which may be attributed to the effect of the interchain
coupling.  As expected, a transition to the long-range ordered state
is found to occur at 1.93 K.

\subsection{NMR spectrum and the Knight shift}
\begin{figure}[t]
\centerline{\psfig{figure=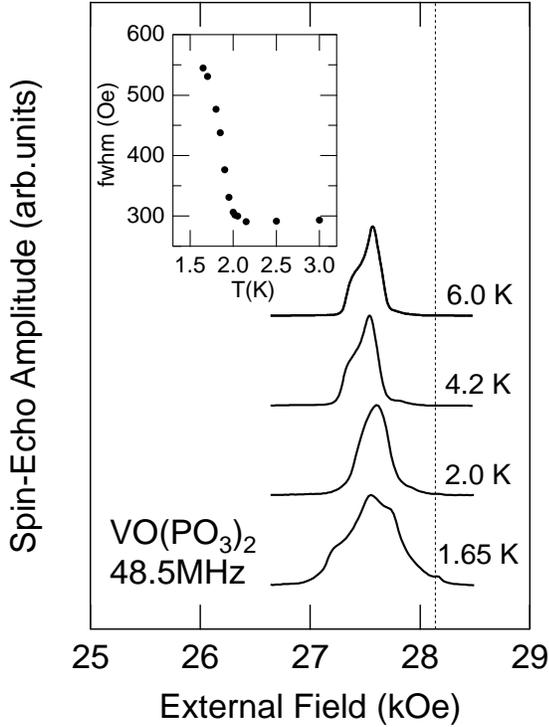,width=9.0cm,angle=0}}
\caption{Temperature variation of the field-swept NMR spectrum of
$^{31}$P in $\alpha$-VO(PO$_3$)$_2$ taken at 48.5 MHz.  The dotted line
indicates a zero-shift position for $^{31}$P. The inset shows the
temperature dependence of FWHM of the spectrum.}
\label{fig:spectra}
\end{figure}

\begin{figure}[t]
\centerline{\psfig{figure=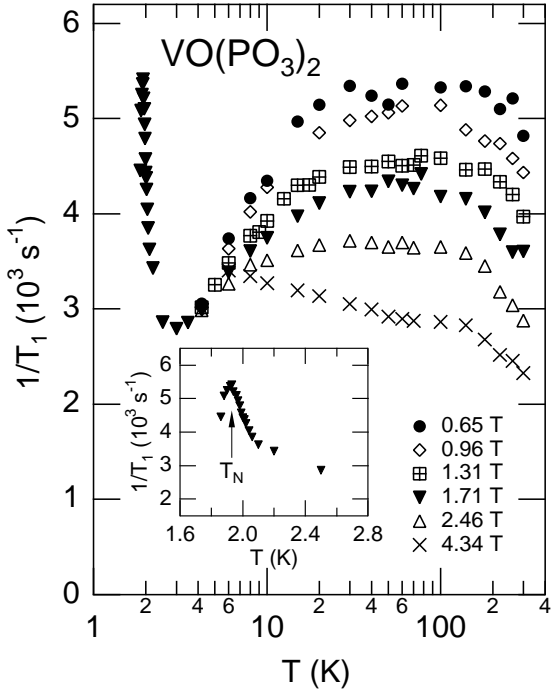,width=9.0cm,angle=0}}
\caption{Temperature dependence of the nuclear spin-lattice relaxation
rate at the P site in $\alpha$-VO(PO$_3$)$_2$ measured at various
magnetic fields.  The inset shows the relaxation rate near $T_{\rm N}$
measured at 1.71 T.}
\label{fig:T1vsT}
\end{figure}

Typical examples of the field-swept $^{31}$P NMR spectrum are shown in
Fig.\ \ref{fig:spectra}.  Above 2.0 K the spectrum exhibits an
asymmetric pattern resulting from an axially-symmetric Knight-shift
tensor.  The spectrum at 2.0 K has some broadening as manifested by
smearing of the fine structure which appears at higher temperatures. 
As the temperature is decreased further, the line becomes broader and
exhibits again a characteristic shape with two shoulders on both sides
of the peak (except the small one near the zero shift which probably
comes from a trace of nonmagnetic impurity phases).  The broadening of
the spectrum signals the onset of long-range magnetic ordering of
V$^{4+}$ spins giving rise to a finite internal magnetic field at the
P site.  The ordering is considered to be antiferromagnetic because
the spectrum is broadened almost symmetrically about its
center-of-gravity position in the paramagnetic state.  As shown in the
inset of Fig.\ \ref{fig:spectra}, the FWHM of the spectrum exhibits a
sudden increase around 2.0 K which gives a rough estimate of the N\'{e}el
temperature $T_{\rm N}$.  Indeed, we determined $T_{\rm N}$ more
precisely from the $T$ dependence of the nuclear spin-lattice
relaxation rate $1/T_1$ at the P site to be 1.93(1) K where $1/T_1$ is
strongly peaked due to critical slowing down of the electronic spins (see
Fig.\ \ref{fig:T1vsT}).

The NMR spectrum below $T_{\rm N}$ does not have a
rectangular shape which we usually observe for nonmagnetic nuclei 
in collinear antiferromagnets.  The observed line shape, however, can be
explained as the one in the two-sublattice antiferromagnet
by taking account of 1) the anisotropy of the hyperfine
coupling resulting in the anisotropic Knight-shift tensor in the
paramagnetic state, 2) incomplete cancellation of the transferred
hyperfine fields from neighboring V$^{4+}$ spins belonging to
different sublattices due to the difference of the interatomic
distances between V and P atoms, and 3) small tipping in the moment
direction from the easy axis under the applied field of which effect
is expressed by the parallel and perpendicular susceptibilities. 
Details of the calculation and the simulation of the spectrum will be
given in the appendix, and we only mention here that as shown in Fig.\
\ref{fig:simulation} the observed spectrum is well reproduced using
physically-reasonable parameters.
\begin{figure}[t]
\centerline{\psfig{figure=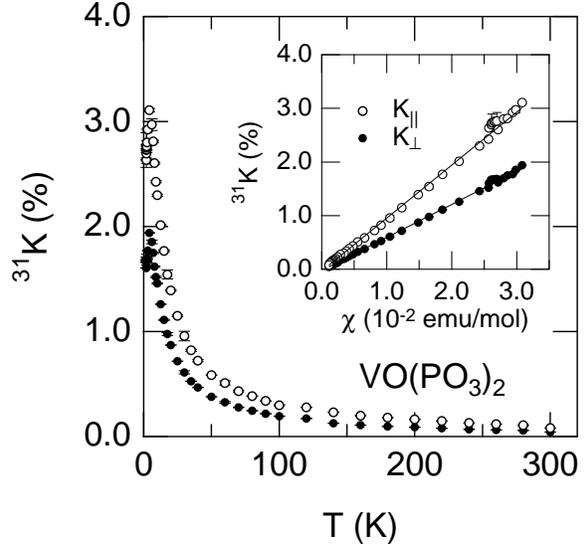,width=9.0cm,angle=0}}
\caption{Temperature dependence of the principal values of the
Knight-shift tensor, $K_{\|}$ (open circles) and $K_{\bot}$ (solid circles), at
the P site in $\alpha$-VO(PO$_3$)$_2$.  The inset shows the $K$-$\chi$
plot for each principal component.  The solid lines in the inset
are the linear fit of the data, the slopes of which yield the
hyperfine coupling constants $A_{\|}$ and $A_{\bot}$.}
\label{fig:KvsT}
\end{figure}

From the values of the intrachain coupling $J$ and the N\'{e}el
temperature $T_{\rm N}$, we can estimate the interchain coupling.  The
interchain coupling $J^{\prime}$ may be evaluated using the
expression\cite{kondo84}
\begin{equation}
    {T_{\rm N} \over J} = \sqrt{z|J^{\prime}| \over 2J}.
\label{eq:interchain}
\end{equation}

\noindent
Substituting $T_{\rm N}=1.93$ K and $J/k_{\rm B}=3.50$ K into eq.\
(\ref{eq:interchain}), we obtain $z|J^{\prime}|/k_{\rm B}=2.13$ K. If we
disregard for simplicity the difference between the second and third
nearest neighbor V-V distances,\cite{fn:dvv} $z=4$ and we get the
``average'' interchain coupling $|J^{\prime}|/k_{\rm B}=0.53$ K (the
sign cannot be specified).  The ratio $|J^{\prime}|/J=0.15$ between
the intra- and interchain exchange couplings is much larger than those
of the other well-known examples of a 1D magnet having $J^{\prime}/J$
of order $10^{-2}$ or less.

A careful analysis of the line shape in the paramagnetic state
enables us an independent determination of the principal values of the
Knight-shift tensor.\cite{abragam61} We plotted in Fig.\
\ref{fig:KvsT} the $T$ dependence of the two independent principal
components of the Knight-shift tensor, $K_{\|}$ and $K_{\bot}$, which
correspond to the shift for the magnetic-field direction parallel and
perpendicular to the symmetry axis at the P site, respectively.  The
Knight-shift tensor depends strongly on temperature and takes a
maximum at 4.2 K as the bulk $\chi$.  As shown in the inset of Fig.\
\ref{fig:KvsT}, both $K_{\|}$ and $K_{\bot}$ scale with the bulk
$\chi$ above and below the susceptibility maximum.  From the linear
slopes of the $K_{\|}$ and $K_{\bot}$ versus $\chi$ plots, we
determined the principal components of the hyperfine tensor at the P
site to be $A_{\|}=5.5(1)$ kOe/$\mu _{\rm B}$ and $A_{\bot}=3.3(1)$
kOe/$\mu _{\rm B}$.  These values yield the isotropic and uniaxial
components of the hyperfine coupling, $A_{\rm
iso}=(A_{\|}+2A_{\bot})/3=4.1(1)$ kOe/$\mu _{\rm B}$ and $A_{\rm
ax}=(A_{\|}-A_{\bot})/3=1.1(1)$ kOe/$\mu _{\rm B}$, respectively.  The
hyperfine coupling at the P site is dominated by an isotropic
transferred hyperfine field from the neighboring V$^{4+}$ spins and
has a small anisotropic component.  $A_{\rm ax}$ is much larger than
is expected from the classical dipolar field of the surrounding
V$^{4+}$ spins ($\sim 0.18$ kOe/$\mu _{\rm B}$) and may be attributed
to polarization of anisotropic $p$ orbitals on the P atom.

\subsection{Nuclear spin-lattice relaxation}
Figure \ref{fig:T1vsT} shows the $T$ dependence of the nuclear
spin-lattice relaxation rate $1/T_1$ at the P site under various
magnetic fields.  The relaxation rate is strongly field dependent
and is larger at lower fields.  The field dependence becomes
weaker as the temperature decreases, and $1/T_1$ becomes almost
field-independent at 4.2 K. As to the $T$ dependence, $1/T_1$ measured
at different fields have some common characteristics. At high
temperatures above about 50 K, $1/T_1$ is only weakly $T$ dependent
and decreases gradually on increasing temperature.  On the other hand,
the relaxation rate decreases rapidly below about 50 K except the rate
at the field $H=4.34$ T showing a modest upturn in that temperature
region.  Note that the decrease of $1/T_1$ at low temperatures is
more pronounced at lower fields.  Below about 2.5 K, $1/T_1$ exhibits
a critical increase toward the long-range magnetic-ordering
temperature as shown in the inset of Fig.~\ref{fig:T1vsT}.  The
maximum of $1/T_1$ is observed at 1.93 K which we determined as the
N\'{e}el temperature of this compound.
\begin{figure}[t]
\centerline{\psfig{figure=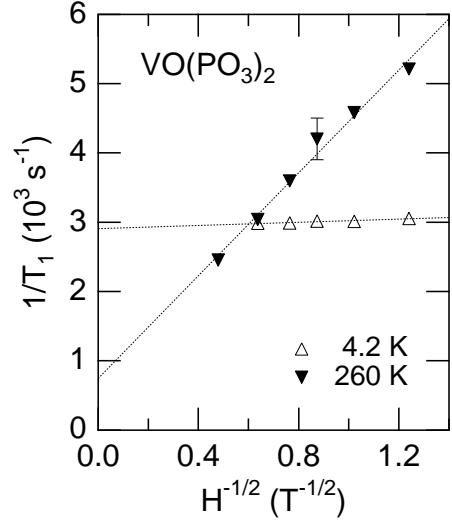,width=7.5cm,angle=0}}
\caption{Nuclear spin-lattice relaxation rate at the P site in
$\alpha$-VO(PO$_3$)$_2$ as a function of $H^{-1/2}$ at two different
temperatures.  The dotted lines are the fit of the data to the formula
$1/T_1=P+QH^{-1/2}$.}
\label{fig:T1vsH}
\end{figure}

The strong magnetic-field ($H$) dependence of $1/T_1$ is likely to come from
spin diffusion characteristic of a low-dimensional Heisenberg spin
system.  In one dimension the spectral density of the spin-spin
correlations diverges as $\omega^{-1/2}$ toward $\omega \rightarrow
0$ which can be probed as $H^{-1/2}$ dependence of $1/T_1$. 
Figure \ref{fig:T1vsH} shows examples of the $H$ dependence of $1/T_1$
at the P site plotted as a function of $H^{-1/2}$.  It is clear that
$1/T_1$ at high temperatures obeys an $H^{-1/2}$ law indicating
diffusive behavior of the spin-spin correlations.  The field
dependence can be fitted to the form
\begin{equation}
    1/T_1 = P + QH^{-1/2}
\label{eq:T1fitting}
\end{equation}

\noindent where $P$ and $Q$ are fitting constants.  While the second
term represents the contribution of spin diffusion near $q=0$, the
first term includes all possible field-independent contributions to
$1/T_1$.  We analyzed the measured $1/T_1$ based on
eq.~(\ref{eq:T1fitting}) down to 4.2 K to parametrize the $T$ and $H$
dependence of $1/T_1$, although the phenomenological theory of spin
diffusion is justified in the limit $T\gg J/k_{\rm B}$.  Physical
meanings of the obtained parameters will be examined in the next
section.  Returning to Fig.\ \ref{fig:T1vsH}, the diffusive
contribution proportional to $H^{-1/2}$ dominates the nuclear-spin
relaxation at 260 K. The slope in $H^{-1/2}$ decreases monotonically
with decreasing temperature while the field-independent part of
$1/T_1$ grows up.  The $H^{-1/2}$ contribution vanishes almost
entirely at 4.2 K, so that $1/T_1$ is governed by field-independent
relaxation processes.  Takigawa {\it et al.} reported a weak $H$
dependence of $1/T_1$ in Sr$_2$CuO$_3$ with a decreasing slope in the
$1/T_1$ vs $H^{-1/2}$ plot on cooling, \cite{takigawa96-2} similar to
the behavior of $1/T_1$ at the P site in $\alpha$-VO(PO$_3$)$_2$ at
low temperatures.
\begin{figure}[t]
\centerline{\psfig{figure=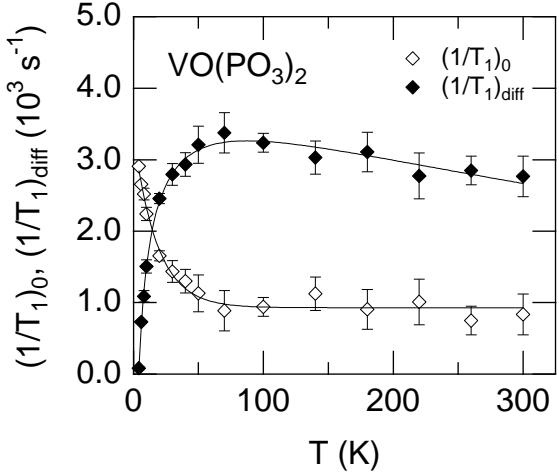,width=9.0cm,angle=0}} 
\caption{Temperature dependence of the field-independent relaxation
rate $(1/T_1)_{\rm 0}$ (open symbols) and the diffusive contribution
$(1/T_1)_{\rm diff}$ at 1.71 T (solid symbols).  The solid lines are
guides to the eyes.}
\label{fig:BGrate}
\end{figure}

The $T$ dependence of the field-independent relaxation rate
$(1/T_1)_{\rm 0}\equiv P$ is shown in Fig.\ \ref{fig:BGrate}. 
$(1/T_1)_{\rm 0}$ is almost $T$ independent with the value of about
$1.0\times 10^{3}$ s$^{-1}$ above 50 K.  Below about
50 K, on the other hand, a substantial increase of $(1/T_1)_{\rm 0}$ was
observed.  This is contrasted with the behavior of $1/T_1$ in that
temperature region where $1/T_1$ decreases rapidly on cooling. 
The low-$T$ decrease of $1/T_1$ therefore results from the decrease of
diffusive contribution which overrides the increasing
contribution of $(1/T_1)_{\rm 0}$.  It is likely that the increase of
$(1/T_1)_{\rm 0}$ from the high-$T$ asymptotic value is caused by an
enhancement of short-range AF correlations within the chain, because
geometry at the P site allows intrachain AF spin fluctuations to
contribute to the $^{31}$P nuclear-spin relaxation.

In Fig.\ \ref{fig:BGrate} we also show for comparison the $T$ dependence of
diffusive contribution $(1/T_1)_{\rm diff} = 1/T_1 - (1/T_1)_{\rm 0}$ at 1.71
T. It is clear that, while govering $1/T_1$ at higher temperatures,
$(1/T_1)_{\rm diff}$ decreases rapidly below about 50 K. The decrease
of $(1/T_1)_{\rm diff}$ synchronizes with the increase of
$(1/T_1)_{\rm 0}$, which strongly suggests some common origin for
them.  As mentioned above, the intrachain short-range AF correlation
is a likely source for such $T$-dependent behavior of the nuclear-spin
relaxation.

\begin{figure}[t]
\centerline{\psfig{figure=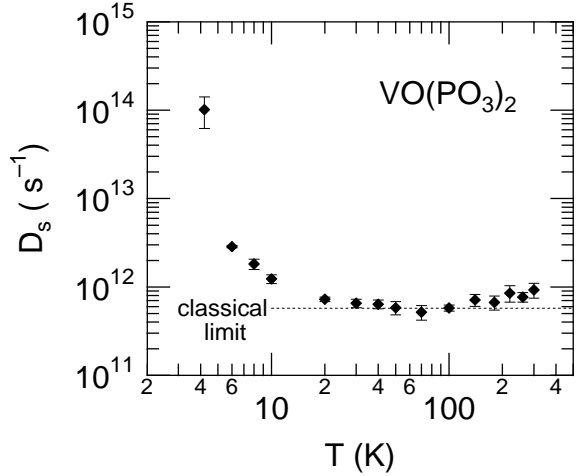,width=9.2cm,angle=0}} 
\caption{Temperature dependence of the spin-diffusion constant $D_s$.  The
dotted line is a value of $D_s$ in the classical limit.}
\label{fig:DSvsT}
\end{figure}

From the slope $Q$ of the $1/T_1$ versus $H^{-1/2}$ plot, we can
estimate the spin-diffusion constant $D_{\rm s}$ which determines the decay of
the spin-spin correlation functions for small $q$ as $\langle S_{q}^{x}
(t) S_{-q}^{x} (0)\rangle \propto \exp(-D_{\rm s}q^2 t) \exp(-i
\omega_{\rm e}t)$ where $\omega_{\rm e}=g\mu_{\rm B}H/\hbar$ is the
electron Larmor frequency.\cite{fn:decay} If the hyperfine coupling is
predominantly isotropic which is the case for the P site in
$\alpha$-VO(PO$_3$)$_2$, the transverse component of the
spectral density $S_{xx}(\omega) = \Sigma_{q} \int_{-\infty}^{\infty} dt e^{i\omega t}
\langle \{S_{q}^{x} (t), S_{-q}^{x} (0)\} \rangle$ dominates the
$H$-dependent part of $1/T_1$.  The nuclear spin-lattice relaxation
rate $(1/T_1)_{\rm diff}$ due to spin diffusion is then given as
\begin{equation}
    \left(\frac{1}{T_1}\right)_{\rm diff}=\frac{A^2}{\hbar^2}\frac{k_{\rm B}T}
    {g^2 \mu_{\rm B}^2}\frac{\chi_{\rm spin}}{\sqrt{2D_{\rm s}\omega_{\rm e}}}
\label{eq:diffusiveT1}
\end{equation}

\noindent
in one dimension. Here $A$ is the hyperfine coupling constant in units of
energy and $\chi_{\rm spin}$ is the spin susceptibility per magnetic atom. 
We determined $D_{\rm s}$ using the values $A_{\rm iso}$ for $A$,
$\chi_{\rm spin} = (\chi-\chi_{0})/N_{\rm A}$ and $g$ determined from
the analysis of the susceptibility and the Knight shift.  The result
is shown in Fig.\ \ref{fig:DSvsT} where $D_{\rm s}$ is plotted as a
function of temperature.  $D_{\rm s}$ is nearly $T$ independent with
the value of $(6-8)\times 10^{11}$ s$^{-1}$ above 20 K. It agrees well
with the classical limit\cite{hone74} $(J/\hbar) \sqrt{2\pi S(S+1)/3}
= 5.7\times 10^{11}$ s$^{-1}$, which suggests that the spin dynamics
is governed by the classical spin diffusion.  On the other hand,
$D_{\rm s}$ becomes unusually large below about 10 K compared with the
classical value.  Such an anomalously large $D_{\rm s}$ may be a
signature that spin diffusion no longer describes the intrinsic
dynamics of the $S=1/2$ HAFC at low temperatures.

\section{Discussion}
\label{sec:discussion}
Although $\alpha$-VO(PO$_3$)$_2$ undergoes a long-range AF transition
at $T_{\rm N}=1.93$ K, the spin dynamics still possesses a
one-dimensional (1D) diffusive character at high temperatures as
manifested by the strong $H^{-1/2}$ dependence of $1/T_1$.  On the one
hand, a weaker field dependence of $1/T_1$ below about 50 K implies
low-energy spin excitations being governed by different physics at
lower temperatures.  The change of the $H$-dependent behavior of
$1/T_1$ around 50 K is nothing but an evidence for crossover between
the two distinct temperature regimes of interest we mentioned in the
Introduction, the paramagnetic regime $T \gg J/k_{\rm B}$ at
temperatures well above the intrachain exchange and the
short-range-ordered (SRO) regime $T \sim J/k_{\rm B}$ with strong
intrachain AF correlations.

In the paramagnetic regime above about 50 K, physical quantities
characterizing the dynamics such as $1/T_1$, $(1/T_1)_{\rm 0}$ and
$D_{\rm s}$ are nearly $T$ independent.  $1/T_1$ is governed by 1D
diffusive spin excitation near $q=0$ giving rise to the $H^{-1/2}$
dependence of $1/T_1$ at the P site.  On the other hand, the
field-independent relaxation rate $(1/T_1)_{\rm 0}$ remains small.  The
spin-diffusion constant $D_{\rm s}$ in this temperature regime is in
good agreement with the value expected in the classical limit, which
suggests that the observed diffusive behavior is an intrinsic property
of the spin system but not an effect of the coupling with the other
degrees of freedom such as phonons.\cite{narozhny96}

It has been argued that the field-independent relaxation rate in the
paramagnetic diffusion regime comes from the longitudinal component of
the spectral density $S_{zz}(\omega)$, of which divergence toward
$\omega \rightarrow 0$ is cutoff by interactions that break the
conservation of the intrachain uniform
magnetization.\cite{hone74,borsa74}.  Here we estimate such a
contribution to $(1/T_1)_{\rm 0}$.  The cutoff effect is taken into
account phenomenologically by multiplying the exponential decay to the
longitudinal diffusive spin correlation function as $\langle S_{q}^{z}
(t) S_{-q}^{z} (0)\rangle \propto \exp(-D_{\rm s}q^2
t)\exp(-\omega_{\rm c}t)$ where $\omega_{\rm c}$ is the cutoff
frequency\cite{takigawa96-1,kikuchi97}.  On the assumption that
the nuclear Larmor frequency is much smaller than $\omega_{\rm c}$,
the relaxation rate $(1/T_1)_{\rm cutoff}$ due to the longitudinal
component with cutoff is given as
\begin{subequations}
\begin{eqnarray}
    \left(\frac{1}{T_1}\right)_{\rm
    cutoff}=\frac{{A^{\prime}}^2}{\hbar^2}\frac{k_{\rm B}T} {g^2 \mu_{\rm
    B}^2}\frac{\chi_{\rm spin}}{\sqrt{D_{\rm s}\omega_{\rm c}}}
\end{eqnarray}
which in the limit $T\rightarrow \infty$ can be written as
\begin{eqnarray}
    \left(\frac{1}{T_1}\right)_{\rm
    cutoff}=\frac{{A^{\prime}}^2}{\hbar^2}\frac{S(S+1)}{3\sqrt{D_{\rm
    s}\omega_{\rm c}}}.
\label{eq:cutoffT1}
\end{eqnarray}
\end{subequations}

\noindent
Here $A^{\prime}$ is the relevant coupling constant which in the
present case is of dipolar origin.  An important cutoff mechanism in
$\alpha$-VO(PO$_3$)$_2$ is the interchain coupling much larger than
the dipolar coupling between electronic spins.  Taking $\omega_{\rm c}
= k_{\rm B}J^{\prime}/\hbar\approx 7 \times 10^{10}$ s$^{-1}$ and
$D_{\rm s} \approx 7 \times 10^{11}$ s$^{-1}$, and using the dipolar
coupling constant ($\sim 0.18$ kOe/$\mu _{\rm B}$) for $A^{\prime}$,
we obtain $(1/T_1)_{\rm cutoff}= 17$ s$^{-1}$ from
eq.~(\ref{eq:cutoffT1}).  This is nearly two orders-of-magnitude
smaller than the observed $(1/T_1)_{\rm 0}$.  Even if we take smaller
$\omega_{\rm c}$ due to the electron dipolar coupling ($\sim$ 0.05 K),
$(1/T_1)_{\rm cutoff}$ is at most 54 s$^{-1}$ and cannot explain the
observed $(1/T_1)_{\rm 0}$.  The above estimate of $(1/T_1)_{\rm
cutoff}$ leads us to conclude that the field-independent relaxation rate
$(1/T_1)_{\rm 0}$ in the paramagentic regime is determined
by the mechanism other than spin diffusion, although the origin 
is not clear at present.  It is noted that the exchange-narrowing
limit of $1/T_1$\cite{moriya56} calculated using parameters for
$\alpha$-VO(PO$_3$)$_2$ as $1/T_{1\infty} = 2.6\times 10^{3}$ s$^{-1}$
cannot explain as well the value of $(1/T_1)_{\rm 0}$ at high temperatures.

Next we discuss the dynamics at low temperatures below about 50 K. An
important observation in this SRO regime is a decreasing contribution
of spin diffusion to $1/T_1$.  Concurrently with a growth of
intrachain AF correlations, diffusive excitations become less
important as a possible channel for nuclear-spin relaxation, and at
temperatures $T \approx J/k_{\rm B}$ where the AF alignment of the
spins along the chain is almost established, the diffusive
contribution becomes essentially absent.  This accompanies an increase of
$D_{\rm s}$ as a fitting variable from the $T$-independent asymptotic
value in the paramagnetic regime.  The apparent increase of $D_{\rm s}$,
however, seems to make no quantitative sense because $D_{\rm s}$ is
estimated using eq.~(\ref{eq:diffusiveT1}) which is justified in the
limit $T\gg J/k_{\rm B}$.  It may rather suggests the classical
spin-diffusion theory becoming inapplicable in the SRO regime: $D_{\rm
s}$ cannot be defined as a physical-meaningful spin-diffusion
constant.  This may result from a qualitative change of the intrinsic
dynamics of the $S=1/2$ HAFC from diffusive to nondiffusive, possibly
propagating ones as approaching $T\approx J/k_{\rm B}$.

It should be noted that the breakdown of the classical spin-diffusion
theory due to SRO does not affect the qualitative behavior of
$(1/T_1)_{\rm 0}$.  Since $1/T_1$ tends to be $H$-independent as the
temperature decreases, it is expected that the result for
$(1/T_1)_{\rm 0}$ does not depend on the precise form of $1/T_1$ as a
function of $H$.  We may therefore conclude that the low-$T$ increase of
$(1/T_1)_{\rm 0}$ is not an artifact but an intrinsic property of the
system.  Although we cannot exclude a possibility that the unidentified
$H$-independent contribution in the paramagnetic regime dominates
$(1/T_1)_{\rm 0}$ in the SRO regime as well, it seems reasonable to
consider that the increase of $(1/T_1)_{\rm 0}$ comes from a growing
contribution of the AF excitation mode because $1/T_1$ exhibits a divergent
behavior toward $T_{\rm N}$.

In summary, the low-frequency spin dynamics in $\alpha$-VO(PO$_3$)$_2$ is
characterized by the two distinct temperature regimes and the crossover
behavior between them.  In the high-$T$ regime ($T\gg J/k_{\rm B}$)
$1/T_1$ is dominated by the $H$-dependent diffusive contribution which
is described quantitatively by the 1D classical spin-diffusion theory.
As the temperature approaches the intrachain coupling strength
$J/k_{\rm B}$, the system goes into a different regime with an almost
$H$-independent $1/T_1$.  If we continue to analyze $1/T_1$ in terms
of the classical spin-diffusion theory, we observe an apparent increase of
the spin-diffusion constant from the classical limit.  It represents
parametrically the decrease of diffusive contribution to $1/T_1$, and
may be interpreted as a sign of crossover to the low-$T$ regime with
nondiffusive, propagating spin excitations.


\appendix
\section{Analysis of the NMR Line Shape in the Antiferromagnetic State}
The NMR line shape observed in the antiferromagnetic (AF) state is
unusual and is to be examined further.  We present in this appendix
the derivation of the resonance condition appropriate for this
specific example and may be applicable if the hyperfine coupling at
the nuclear site under consideration has uniaxial symmetry.  We also
give the calculation of the line shape using the derived resonance
condition.  From comparison of the calculated spectrum with the
observed one, it is shown that the observed line shape below
$T_{\rm N}$ is in consistency with collinear antiferromagnetic structure
having two sublattices.  The point of derivation is to take account of an
anisotropy of the hyperfine coupling, incomplete cancellation of the
transferred hyperfine fields from neighboring magnetic ions, and
field-induced canting of sublattice magnetizations.

The resonance frequency $\omega$ for nuclei with the
nuclear gyromagnetic ratio $\gamma$ is generally given as
\begin{equation}
    \omega / \gamma = |{\mib H}_{\rm 0}+{\mib H}_{\rm n}|.
\label{eq:condition}
\end{equation}

\noindent
Here ${\mib H}_{\rm 0}$ and ${\mib H}_{\rm n}$ are external and
internal magnetic fields, respectively.  Neglecting the classical
dipolar field for the time being, ${\mib H}_{\rm n}$ is due to the
hyperfine field and is given as
\begin{equation}
    {\mib H}_{\rm n} = \sum_{j} \widetilde{\mib A}_{ij} \cdot {\mib S}_{j}
\label{eq:Hn}
\end{equation}

\noindent
where ${\mib S}_{j}$ is the electronic spin at the $j$-th site and
$\widetilde{\mib A}_{ij}$ is the hyperfine tensor between the sites
$i$ and $j$.  ${\mib S}_{j}$ fluctuates in time and hence
${\mib H}_{\rm n}$ in eq.~(\ref{eq:condition}) should be the
time average because we are dealing with the resonance frequency.  For
simplicity of notation we omit the brackets $\langle~\rangle$ to show
explicitly the time-averaging procedure and use ${\mib S}_{j}$ and
${\mib H}_{\rm n}$ as the time-averaged quantities in the following.

In $\alpha$-VO(PO$_3$)$_2$, there are two neighboring V atoms in the
distances 3.292 {\AA} and 3.392 {\AA} around the P site and the others
are relatively far apart ($\geq 4.613$ {\AA}).  Because a transferred
hyperfine coupling is generally short-ranged, we may consider a
contribution of these two V atoms to ${\mib H}_{\rm n}$ at the P site. 
Labelling these V sites as V(1) and V(2), and similarly the electronic
spins as ${\mib S}_{1}$ and ${\mib S}_{2}$, we can rewrite eq. 
(\ref{eq:Hn}) as
\begin{equation}
    {\mib H}_{\rm n} = \widetilde{\mib A}_{1} \cdot {\mib S}_{1} +
    \widetilde{\mib A}_{2} \cdot {\mib S}_{2}
\label{eq:Hn_P}
\end{equation}

\noindent
for the P site in $\alpha$-VO(PO$_3$)$_2$.  Here $\widetilde{\mib A}_{1}$ and
$\widetilde{\mib A}_{2}$ are the hyperfine tensors from V(1) and V(2)
sites, respectively.  Since V(1) and V(2) belong to the different AF
chains, ${\mib S}_{1}$ and ${\mib S}_{2}$ can in principle be either
parallel or antiparallel in zero external field.

If ${\mib S}_{1}$ and ${\mib S}_{2}$ are parallel, i.e., V(1)
and V(2) belong to the same sublattice, ${\mib S}_{1} = {\mib S}_{2}$
and hence ${\mib H}_{\rm n} = (\widetilde{\mib A}_{1} +
\widetilde{\mib A}_{2}) \cdot {\mib S}_{1}$.  Because the sum
$\widetilde{\mib A}_{1} + \widetilde{\mib A}_{2}$ corresponds to the
hyperfine tensor in the paramagnetic state, there should be a
transferred field of significant magnitude (a few kOe) at the P site. 
We expect in that case a usual rectangular shape of the spectrum which
contradicts with the observation.  We therefore assume antiparallel
alignment of ${\mib S}_{1}$ and ${\mib S}_{2}$ under zero external
field which means that V(1) and V(2) belong to different sublattices.
\begin{figure}[t]
\centerline{\psfig{figure=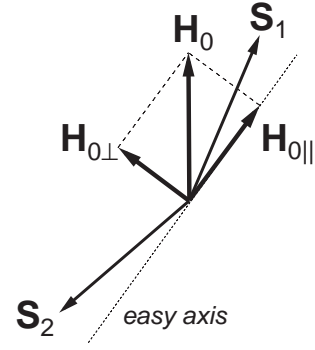,width=4.0cm,angle=0}}
\medskip 
\caption{Schematic view of the canting due to the external field
${\mib H}_{0}$ of the electronic spins ${\mib S}_{1}$ and ${\mib
S}_{2}$.  The dotted line indicates the direction of the easy
axis.  Decomposition of ${\mib H}_{0}$ to the components parallel and
perpendicular to the easy axis, ${\mib H}_{0\|}$ and ${\mib
H}_{0\bot}$, respectively, is also shown.}
\label{fig:canting}
\end{figure}

Under nonzero external field, the sublattice moment changes its direction
and magnitude slightly from the zero-field values.  If the external field is
smaller than the spin-flop field, the effect may be expressed by the
parallel and perpendicular susceptibilities, $\chi_{\|}$ and
$\chi_{\bot}$, in the AF state.  It is then convenient to decompose
${\mib S}_{1}$ and ${\mib S}_{2}$ into components parallel and
perpendicular to the easy axis for the sublattice moment as ${\mib
S}_{j} = {\mib S}_{j\|} +{\mib S}_{j\bot}~(j=1,2)$ and similarly the
external field as ${\mib H}_{0} = {\mib H}_{0\|} +{\mib H}_{0\bot}$
(Fig.\ \ref{fig:canting}).  The following relations hold between the
susceptibilities and the parallel and perpendicular components of
${\mib S}_{1}$, ${\mib S}_{2}$ and ${\mib H}_{0}$;
\begin{subequations}
\begin{equation}
    {\mib S}_{1\|} + {\mib S}_{2\|} = 2 \chi_{\|} {\mib H}_{0\|},
\end{equation}
\begin{equation}
    {\mib S}_{1\bot} + {\mib S}_{2\bot} = 2\chi_{\bot} {\mib H}_{0\bot}.
\end{equation}
\end{subequations}

\noindent
Here $\chi_{\|}$ and $\chi_{\bot}$ are defined as the susceptibilities
per magnetic atom.  Defining $\widetilde{\mib A} = \widetilde{\mib A}_{1} +
\widetilde{\mib A}_{2}$ and $\widetilde{\mib a} = \widetilde{\mib A}_{1} -
\widetilde{\mib A}_{2}$, eq.~(\ref{eq:Hn_P}) can be written as
\begin{eqnarray}
\label{eq:Hn_P_1}
    {\mib H}_{\rm n} &=& \chi_{\bot} \widetilde{\mib A} \cdot
    {\mib H}_{0\bot} + \chi_{\|} \widetilde{\mib A} \cdot {\mib H}_{0\|} 
    \nonumber \\ &+& 
    \frac{1}{2} [\widetilde{\mib a} \cdot ({\mib S}_{1\bot} - {\mib
    S}_{2\bot}) + \widetilde{\mib a} \cdot ({\mib S}_{1\|} - {\mib
    S}_{2\|})] \\
    &\approx& \chi_{\bot} \widetilde{\mib A} \cdot {\mib H}_{0\bot} +
    \chi_{\|} \widetilde{\mib A} \cdot {\mib H}_{0\|} + \widetilde{\mib a}
    \cdot {\mib S}_{1\|}.
\label{eq:Hn_P_2}
\end{eqnarray}

\noindent
On going from (\ref{eq:Hn_P_1}) to (\ref{eq:Hn_P_2}) we used the
approximations ${\mib S}_{1\|} \approx -{\mib S}_{2\|}$ and ${\mib
S}_{1\bot} \approx {\mib S}_{2\bot}$.  The first and second terms in
eq.~(\ref{eq:Hn_P_2}) correspond to the hyperfine field resulting from
canting of ${\mib S}_{1}$ and ${\mib S}_{2}$ from the easy axis under
nonzero external field.  The third term represents the hyperfine field
due to the difference between $\widetilde{\mib A}_{1}$ and
$\widetilde{\mib A}_{2}$.  Putting $\widetilde{\mib a} \cdot {\mib
S}_{1\|} = {\mib H}_{\rm n0}$ in eq.~(\ref{eq:Hn_P_2}) gives a more
convenient form for ${\mib H}_{\rm n}$;
\begin{equation}
    {\mib H}_{\rm n} = \chi_{\bot} \widetilde{\mib A} \cdot
    {\mib H}_{0\bot} + \chi_{\|} \widetilde{\mib A} \cdot {\mib H}_{0\|} +
    {\mib H}_{\rm n0}.
\label{eq:Hn_P_3}
\end{equation}

\noindent
As has been pointed out, ${\mib H}_{\rm n0}$ comes from uncancellation of the
hyperfine fields from the two neighboring V atoms.

\begin{figure}[t]
\centerline{\psfig{figure=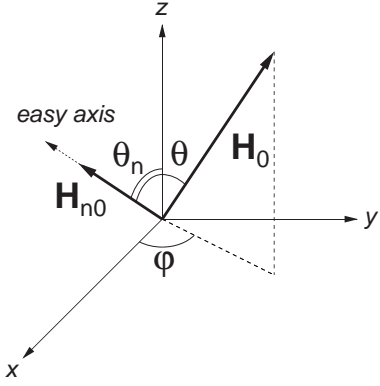,width=5.0cm,angle=0}}
\medskip 
\caption{Principal frame used in the calculation of the resonance
frequency.  The dotted line with an arrow indicates the direction of the
easy axis.  The easy axis is taken to be in the $zx$ plane.}
\label{fig:frame}
\end{figure}

In deriving the expression of the resonance condition from which the
line shape is calculated, we have to take some orthogonal frame and
specify the directions of ${\mib H}_{0}$ and ${\mib H}_{\rm n0}$.  It
is reasonable to take the principal frame of the hyperfine tensor
$\widetilde{\mib A}$ at the P site in which $\widetilde{\mib A}$ is
diagonalized and the principal values are known from the experiment. 
Taking the $z$ axis as the unique axis of the tensor $\widetilde{\mib
A}$ having axial symmetry, we denote the directions of the easy axis
and the external field ${\mib H}_{0}$ by the polar and azimuth angles
as shown in Fig.\ \ref{fig:frame}.  The easy axis can be taken to lie
in the $zx$ plane so that $\varphi_{\rm n} = 0$ without loss of
generality because of the axial symmetry of $\widetilde{\mib A}$.  For
the field ${\mib H}_{\rm n0}$ we assume that it is in the direction of
the easy axis for simplicity.  This is equivalent to neglect the
anisotropy of $\widetilde{\mib a}$.  It is noted that with this
assumption, the classical dipolar field may be included in ${\mib
H}_{\rm n0}$ so that ${\mib H}_{\rm n0}$ is regarded as the sum of
classical dipolar and hyperfine fields, because the dipolar field is a
likely source for the anisotropy field in the $S=1/2$ system and may
be taken to be parallel to the easy axis.

Our aim is to express the resonance frequency $\omega$ as a function of
the angles $\theta$, $\varphi$ and $\theta_{\rm n}$, and the field
strength $H_{0} = |{\mib H}_{0}|$ and $H_{\rm n0} = |{\mib H}_{\rm
n0}|$.  The angles $\theta$ and $\varphi$ distribute randomly in the
polycrystalline sample, so that a specific line shape called powder
pattern is observed.  In the principal frame of $\widetilde{\mib A}$,
${\mib H}_{\rm 0}$ and ${\mib H}_{\rm n}$ are written explicitly as
follows;
\begin{subequations}
\begin{equation}
    \widetilde{\mib A} = \pmatrix{A_{x} &  &  \cr  & A_{y} &  \cr  &
    & A_{z}\cr}= \pmatrix{A_{\bot} & & \cr & A_{\bot} & \cr & & A_{\|}
    \cr},
\end{equation}
\begin{equation}
    {\mib H}_{0} = \pmatrix{b_{x} H_{0} \cr b_{y} H_{0} \cr b_{z} H_{0}
    \cr} = \pmatrix{H_{0} \sin \theta \cos \varphi \cr H_{0} \sin \theta
    \sin \varphi \cr H_{0} \cos \theta \cr},
\end{equation}
\begin{equation}
   {\mib H}_{\rm n0} = \pmatrix{c_{x} H_{\rm n0} \cr 0 \cr c_{z} H_{\rm
   n0} \cr} = \pmatrix{H_{\rm n0} \sin \theta_{\rm n} \cr 0 \cr H_{\rm n0}
   \cos \theta_{\rm n} \cr}.
\end{equation}
\label{eq:components}
\end{subequations}

\noindent
$A_{\|}$ and $A_{\bot}$ are the principal values of $\widetilde{\mib
A}$ which can be determined from the $K$-$\chi$ plots in the paramagnetic
state.  ${\mib H}_{0\|}$ is just the projection of ${\mib H}_{0}$ onto
${\mib H}_{\rm n0}$, and hence ${\mib H}_{0\|}$ and ${\mib H}_{0\bot}$
can also be written down by components.  Substituting
eq.~(\ref{eq:Hn_P_3}) into eq.~(\ref{eq:condition}) and writing
explicitly the components in the principal frame of $\widetilde{\mib
A}$ using eqs.~(\ref{eq:components}), we get after lengthy
calculations
\begin{subequations}
\begin{eqnarray}
    \omega^2 = {\omega_x}^2 + {\omega_y}^2 + {\omega_z}^2
\end{eqnarray} 
with
\begin{eqnarray}
    \omega_x/\gamma &=& [\{1 + A_{\bot} ({c_x}^2 \chi_{\|} + {c_z}^2
    \chi_{\bot})\} b_x \nonumber \\  &&+ A_{\bot} (\chi_{\|} - \chi_{\bot}) c_x
    c_z b_z] H_{0} + c_x H_{\rm n0}, \nonumber \\
    \omega_y/\gamma &=& (1 + A_{\bot} \chi_{\bot}) b_y H_{0}, \\
    \omega_z/\gamma &=& [\{1 + A_{\bot} ({c_z}^2 \chi_{\|} + {c_x}^2
    \chi_{\bot})\} b_z \nonumber \\ &&+ A_{\|} (\chi_{\|} - \chi_{\bot}) c_x c_z
    b_x] H_{0} + c_z H_{\rm n0}.  \nonumber
\end{eqnarray}
\label{eq:condition-2}
\end{subequations}

\noindent
Note that $\chi_{\|}~(\chi_{\bot})$ is not the susceptibility for the
$z~(x,y)$ direction(s) in the principal frame of $\widetilde{\mib A}$ but
the one parallel (perpendicular) to the easy axis.  Neglecting the
terms of $O(A^2 \chi^2)$, and using the relations ${b_x}^2 + {b_y}^2 +
{b_z}^2 = 1$ and ${c_x}^2 + {c_z}^2 = 1$, eq.~(\ref{eq:condition-2})
can be rewritten as
\begin{eqnarray}
\label{eq:condition-3}
    \omega^2 / \gamma^2  &=& {H_{0}}^2 + {H_{\rm n0}}^2 + 2 
    (b_x c_x + b_z c_z) H_{0} H_{\rm n0} \nonumber \\ &+& 2 [A_{\bot} ({c_x}^2
    \chi_{\|} + {c_z}^2 \chi_{\bot}) {b_x}^2 + A_{\bot} \chi_{\bot}
    {b_y}^2 \nonumber \\ &&+ A_{\|} ({c_z}^2 \chi_{\|} + {c_x}^2
    \chi_{\bot}) {b_z}^2 \nonumber \\ &&+ (A_{\|} + A_{\bot}) (\chi_{\|} -
    \chi_{\bot}) b_x b_z c_x c_z ] H_{0}^2 \nonumber \\ &+& 2 [A_{\bot}
    ({c_x}^2 \chi_{\|} + {c_z}^2 \chi_{\bot}) b_x c_x \\ &&+ A_{\|}
    ({c_z}^2 \chi_{\|} + {c_x}^2 \chi_{\bot}) b_z c_z \nonumber \\ &&+
    A_{\|} (\chi_{\|} - \chi_{\bot}) {c_z}^2\, b_x c_x \nonumber \\ && +
    A_{\bot} (\chi_{\|} - \chi_{\bot}) {c_x}^2\, b_z c_z] H_{0} H_{\rm
    n0}.\nonumber
\end{eqnarray}

\noindent
Equation~(\ref{eq:condition-3}) is valid for arbitrary strength of $H_{\rm
n0}$ as far as the conditions $A\chi \ll 1$ are satisfied.

If $H_{\rm n0}$ is much larger than $A\chi$,
eq.~(\ref{eq:condition-3}) may be simplified as
\begin{subequations}
\begin{equation}
    \omega^2 / \gamma^2 = {H_{0}}^2 + {H_{\rm n0}}^2 + 2 (b_x c_x + b_z c_z)
    H_{0} H_{\rm n0}
\end{equation}

\noindent
which, by denoting the angle between ${\mib H}_{0}$ and ${\mib H}_{\rm n0}$
as $\beta$, reduces to
\begin{equation}
    \omega^2 / \gamma^2 = {H_{0}}^2 + {H_{\rm n0}}^2 + 2 H_{0} H_{\rm
    n0} \cos \beta.
\end{equation}
\end{subequations}

\noindent
Random distribution of $\beta$ in the polycrystalline sample yields a
usual rectangular shape of the spectrum.\cite{kikuchi00}

If $H_{\rm n0}$ is much smaller than $H_{\rm 0}$ but is comparable
with $A\chi$ which is the case here, we may neglect in
eq.~(\ref{eq:condition-3}) the terms proportional to ${H_{\rm n0}}^2$
or $A\chi H_{\rm n0}$.  Hence we arrive at the final expression of the
resonance frequency;
\begin{eqnarray}
    \omega / \gamma &=& H_{0} + [A_{\bot} ({c_x}^2 \chi_{\|} +
    {c_z}^2 \chi_{\bot}) {b_x}^2 + A_{\bot} \chi_{\bot} {b_y}^2 \nonumber \\ &&+
    A_{\|} ({c_z}^2 \chi_{\|} + {c_x}^2 \chi_{\bot}) {b_z}^2 \\
    &&+ (A_{\|} + A_{\bot}) (\chi_{\|} - \chi_{\bot}) b_x b_z c_x c_z ]
    H_{0}\nonumber \\ &&+ (b_x c_x + b_z c_z) H_{\rm n0},\nonumber
\end{eqnarray}

\noindent
or writing explicitly the dependence on the angles as
\begin{eqnarray}
\label{eq:condition-4}
    \omega / \gamma  &=& H_{0} + [A_{\bot} (\chi_{\|} \sin^2 \theta_{\rm n} +
    \chi_{\bot} \cos^2 \theta_{\rm n}) \sin^2 \theta \cos^2 \varphi
    \nonumber \\ && + A_{\bot} \chi_{\bot} \sin^2 \theta \sin^2 \varphi
    \nonumber \\ &&+ A_{\|} (\chi_{\|} \cos^2 \theta_{\rm n} + \chi_{\bot}
    \sin^2 \theta_{\rm n}) \cos^2 \theta \\ && + \frac{1}{4} (A_{\|} +
    A_{\bot}) (\chi_{\|} - \chi_{\bot}) \sin 2\theta_{\rm n} \sin 2\theta
    \cos \varphi ] H_{0} \nonumber \\
    &&+ (\sin \theta_{\rm n} \sin \theta \cos \varphi + \cos \theta_{\rm n}
    \cos \theta ) H_{\rm n0}.\nonumber
\end{eqnarray}

\begin{figure}[t]
\centerline{\psfig{figure=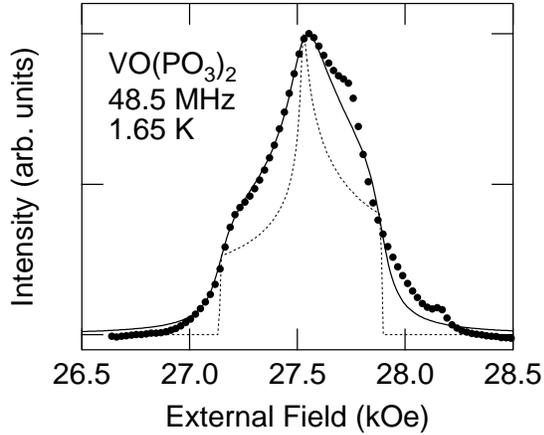,width=9.0cm,angle=0}}
\medskip 
\caption{Comparison of the calculated powder patterns with the observed
spectrum at 1.65 K. The solid circles represent the experimental data.  The
dashed line is the calculated spectrum $f(H_{0})$ with no inhomogeneous
broadening.  The parameters used in the calculation are $\omega =
48.5$ MHz, $A_{\|} = 5.5$ kOe/$\mu _{\rm B}$, $A_{\bot} = 3.3$
kOe/$\mu _{\rm B}$, $H_{\rm n0} = 310$ Oe, $\theta_{\rm n} =
33^{\circ}$, $\chi_{\|} = 0.026$ emu/mol and $\chi_{\bot} = 0.031$
emu/mol.  The spectrum shown by the solid line is calculated by taking
account of inhomogeneous distribution of the internal field $H_{\rm n0}$.  The
effect is introduced by convoluting lorentzian with the FWHM of 55
Oe.}
\label{fig:simulation}
\end{figure}

For the field-sweep measurement the resonance frequency
$\omega$ is fixed while the external field $H_{0}$ is varied.  It is
therefore necessary to solve eq.~(\ref{eq:condition-4}) with respect
to $H_{0}$, the result of which is not given here explicitly because it
is rather straightforward.  Then we calculate the spectrum $f(H_{0})
\propto \Delta N/\Delta H_{0}$ by counting the number of nuclei
$\Delta N$ having the resonance field between $H_{0}$ and $H_{0}
+\Delta H_{0}$. More than million different $(\theta, \varphi)$
points representing the random distribution of $H_{0}$ were taken to
calculate the resonance field.  The unknowns $H_{\rm n0}$,
$\theta_{\rm n}$, $\chi_{\|}$ and $\chi_{\bot}$ in
(\ref{eq:condition-4}) are treated as parameters to reproduce the
observed spectrum.  As shown in Fig.\ \ref{fig:simulation}, an
appropriate choice of the parameter values reproduces the observed
two-shoulder structure of the spectrum in the AF state.  We may
therefore conclude that the magnetic ordering in $\alpha$-VO(PO$_3$)$_2$ is
not unusual but is rather conventional.





\begin{thebibliography}{99}
\bibitem{fn:dvv} The second and third nearest neighbor V atoms are in
the distances of 5.223 {\AA} and 6.092 {\AA} at room temperature.
\bibitem{fn:decay} The power law decay of the spin autocorrelation
function $\langle S_{i}^{x} (t) S_{i}^{x} (0)\rangle \propto t^{-1/2}$
(eq.~(\ref{eq:diffusion})) can be derived by integrating $\langle S_{q}^{x} (t)
S_{-q}^{x} (0)\rangle$ over $q$ in one dimensional space.
\bibitem{fn:brackets}  ${\mib S}_{j}$ fluctuates in time and hence
${\mib H}_{\rm n}$ in eq.~(\ref{eq:condition}) should be the
time average because we are dealing with the resonance frequency.  For
simplicity of notation we omit the brackets $\langle~\rangle$ to show
explicitly the time-averaging procedure and use ${\mib S}_{j}$ and
${\mib H}_{\rm n}$ as the time-averaged quantities.
\end{thebibliography}
\end{document}